\documentclass[a4paper,11pt]{article}
\usepackage{jinstpub} 
\usepackage{lineno}
\usepackage{upgreek}

\title{\boldmath Synchronous and asynchronous Data Quality Control of the  ALICE Inner Tracking System in the LHC Run 3}

\collaboration[c]{on behalf of the ALICE Collaboration}

\author{Svetlana Kushpil}
\affiliation{Nuclear Physics Institute of the CAS,\\
Husinec - Řež, Hlavni str. 130, 250 68 Řež, Czech Republic}

\emailAdd{skushpil@ujf.cas.cz}

\abstract{The Inner Tracking System (ITS) of the ALICE experiment at the CERN Large Hadron Collider (LHC) is the
 largest Monolithic Active Pixel Sensor technology application in high-energy physics.
The upgraded version of the tracking system, called ITS2, consists of seven concentric layers of ALPIDE monolithic active pixel sensors produced in the 180 nm CMOS process, covering a total sensitive area of about 10 m${}^2$.
The ALPIDE sensor features a pixel pitch of 27 $\upmu$m $\times$ 29 $\upmu$m and a position resolution of about 5 $\upmu$m.
The very low material budget, 0.36\% $X_{0}$/layer for the three innermost layers and  1.10\% $X_{0}$/layer for the outer layers, in combination with the small radial distance of only 23 mm from the beam, leads to an excellent impact parameter resolution at low transverse momentum. This makes the detector well suited for experimentally challeging physics measurements such as the reconstruction of low transverse momentum heavy-flavor particles in the heavy-ion collision environment.
This contribution provides an overview of the ITS2 data Quality Control system (QC),
a framework designed to synchronously monitor the detector operating parameters and provide asynchronous reconstruction of the collected data, with the goal of guaranteeing a stable and efficient data taking.
The monitoring for fake-hit rate, front-end electronics status, data integrity, cluster and track distributions, 
are presented, together with an overview of the ITS2 performance during the recent Run 3 pp and Pb--Pb data taking campaigns, as extracted from the QC asynchronous reconstruction.
}


\arxivnumber{1234.56789} 

\begin{document}
\maketitle
\flushbottom
\section{Introduction}

The Inner Tracking System (ITS2) \cite{ITS2,ALICEupgrades} of the ALICE experiment  at the CERN Large Hadron Collider (LHC) is the largest silicon tracker based on Monolithic Active Pixel Sensor (MAPS) technology in high-energy physics. The detector consists of  seven concentric layers of MAPS  sensors called ALPIDE  \cite{ALPIDE}  produced in a 180 nm CMOS process, covering a total sensitive area of about 10 m${}^2$.  Multiple ALPIDEs are assembled as the innermost three layers (Inner Barrel, IB) and the four outer layers (Outer Barrel, OB).
The ITS2 has about $12.5 \times 10^{9}$  pixels with a pixel pitch of 27 $\upmu$m $\times$ 29 $\upmu$m. The very low material budget, 0.36\% $X_{0}$/layer for the three innermost layers and 1.10\% $X_{0}$/layer for the outer layers, in combination with the small radial distance of only 23 mm from the beam,  ensures an excellent impact parameter resolution at low transverse momentum. This makes the detector well suited for experimentally challenging physics measurements such as the reconstruction of low transverse momentum heavy-flavor particles in heavy-ion collisions.

\section{ Data Quality Control System}

In Run 3, ALICE performs data-taking in a continuous readout mode. The recorded data is synchronously reconstructed.  The data quality of ITS2 is monitored during the synchronous reconstruction using the ALICE QC framework \cite{Operation}.   The synchronous processing is realized by 13 ITS2 First Level Processors (FLPs) and 340 Event Processing Nodes (EPNs) shared by all ALICE detectors. 
The characteristics of the FLPs and EPNs and the strategy to process data for the QC tasks are summarized in TDR \cite{FLP_EPN}. Checks based on the raw data from individual segments of the detector are run at the level of the FLP. These QC checks comprise the integrity of the data received from the front-end electronics and monitor the detector occupancy. At the level of the EPN, the data from all detectors for a given time segment 
is available, enabling checks based on information obtained in the synchronous reconstruction as well as those that require information from the full detector. 
The time in-between fills of the LHC is used for calibration measurements, which range from a charge-threshold scan for the ITS2 threshold, threshold tuning, and noisy pixel masking. The QC tasks monitor decoding errors, dead-chip maps, clustering and tracking performance, threshold setting, and noisy pixels. During the synchronous phase, time-dependent acceptance maps for ITS2 are extracted and later used as input for Monte Carlo simulations.  The reconstructed data are then stored on a disk for later asynchronous processing.
The asynchronous processing serves to incorporate improved calibration and is optimized for physics performance. 

In the synchronous phase, the QC runs seven tasks  on a subset of data to perform online monitoring of
 1) data integrity check of all events,  2)  detector occupancy, 3) cluster size and cluster topology,  4) track multiplicity and angular distribution,  5) extraction of noisy pixels for offline noise masks, 6) the thresholds obtained during calibration scans  and  dead pixels, and 7) availability of a chip or high-speed transmission line in the data stream.  
Finally, there is an offline post-processing framework that plots the trends of data obtained during the asynchronous processing.

\section{Comprehensive investigations to achieve the optimal detector performance}

Detector data quality and performance are monitored
   in real time relying on the synchronous QC and are
   studied in more detail in the asynchronous data reconstruction.
 Figure 1 shows an example of the QC on-line monitoring quality assessment plots for 24/7 shifts during a single run of Pb–Pb data-taking. The top part presents a general occupancy plot for a run tagged as ``GOOD''. Fake-hit rate (FHR) is obtained as the number of fake hits normalized per one event and pixel. FHR is measured during data-taking in absence of LHC beams, while in case of actual collisions, the same plot gives us hit occupancy. The plot provides information about the maximal occupancy per ITS stave. Each stave is depicted as a triangle. 
 The bottom part shows chip status plots that monitor
the  fraction of time when chips do not register any hits. The information is provided per high-speed transmission line grouped by detector layers. The horizontal axis shows a time span corresponding to 15 minutes of Pb–Pb data-taking. One QC cycle takes 30 seconds. Bins highlighted with yellow mark the ALPIDE chips that in a given QC cycle  did not send data for at least one readout frame 
(around 15 $\upmu$s).  
The thicker lines corresponding to the full stave not sending data. ITS2 have auto-recovery system which detects problematic lanes and performs recovery on the level of staves which longs (from 10 to 30 seconds) after which lane return to the data-taking. Continuous lines correspond to the dead chips which are not possible to recover with auto-recovery (fraction is ~1$\%$ in OB, while IB does not have such chips at all).
Based on this plot one can identify problems with readout lanes and get an overview of permanently not working chips.

\begin{figure}[htbp]
\centering
\includegraphics[width=.5\textwidth]{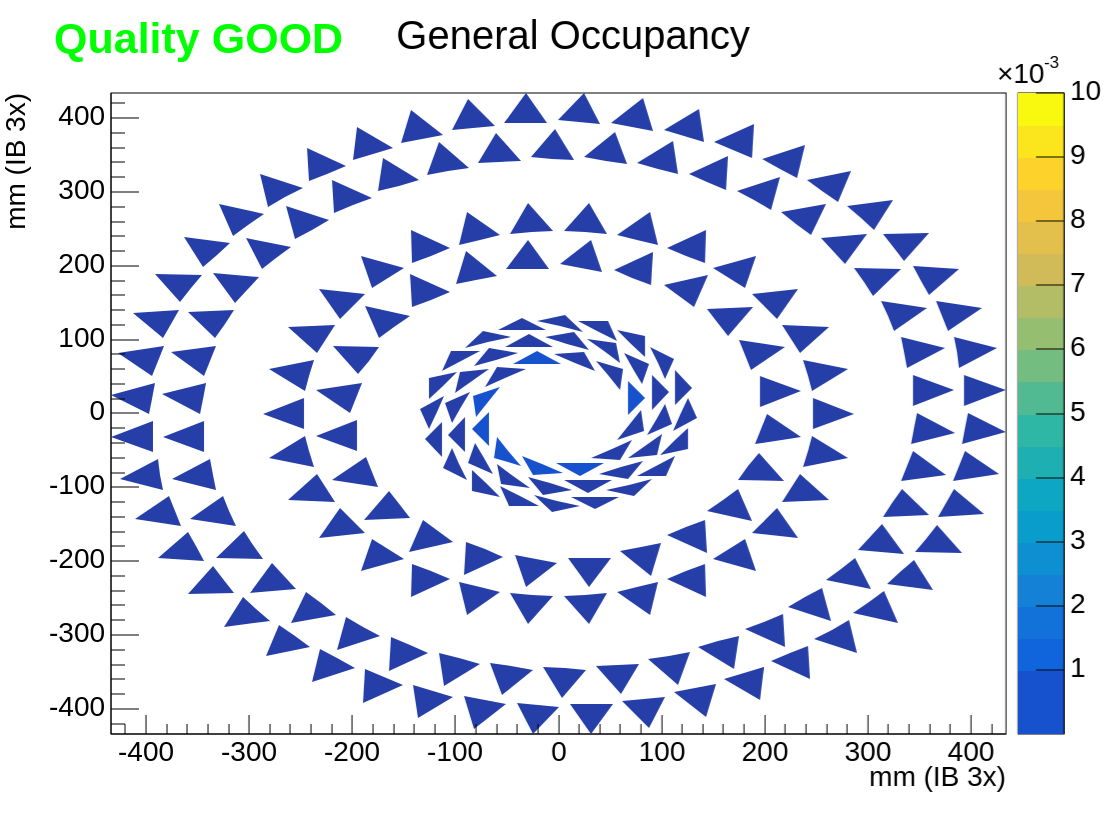}
\qquad
\includegraphics[width=1.\textwidth]{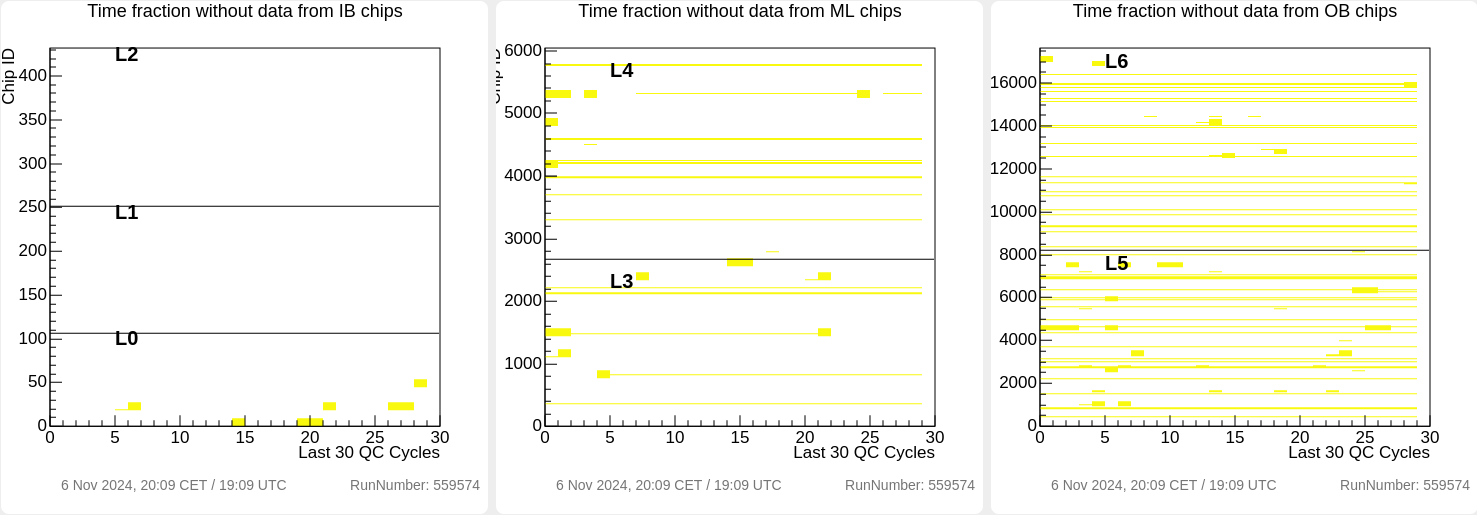}
\caption{QC plots from on-line monitoring\label{fig:i}}
\end{figure}

ITS2 data reconstruction is illustrated by the comparison of synchronous  and asynchronous data reconstruction in the left and right panels of Figure 2 and Figure 3. The plots illustrate ITS2 performance during pp collisions at $\sqrt{s}=13.6$ TeV with a visible interaction rate of 500 kHz and 202 kHz framing rate. 
The observed differences between the left and right panels
are due to different reconstruction algorithms and selection
criteria which are applied during the synchronous and asynchronous
reconstruction. In addition, during the synchronous reconstruction, 
the QC processes only only 1\% of data whereas during the asynchronous reconstruction one analyzes the full dataset.

\begin{figure}[htbp]
\centering
\includegraphics[width=.45\textwidth]{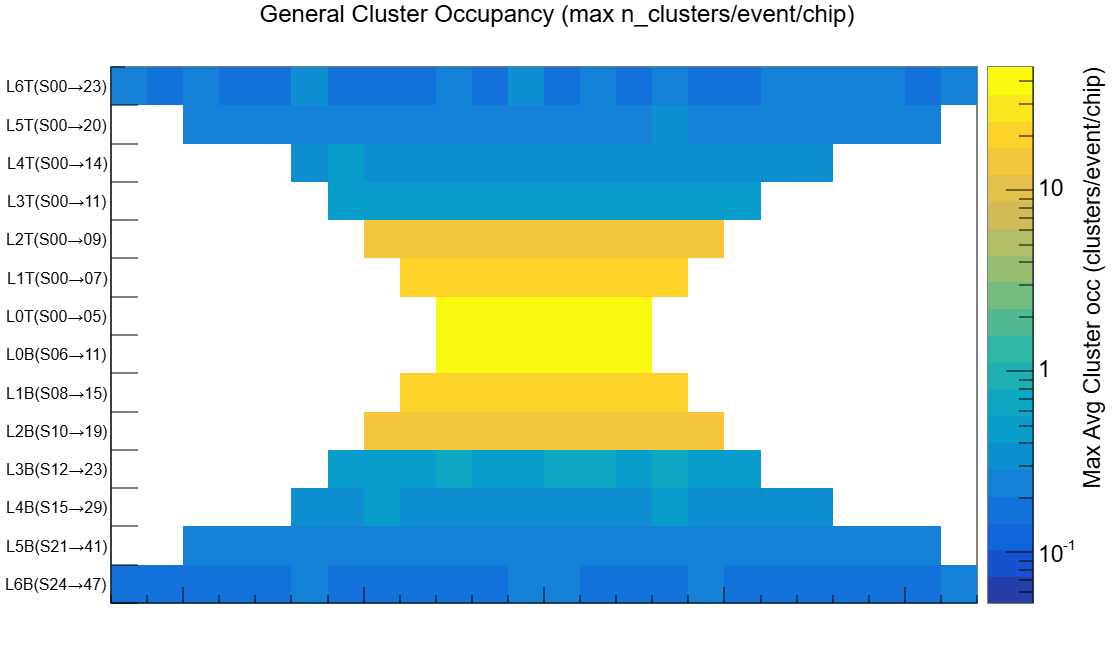}
\qquad
\includegraphics[width=.45\textwidth]{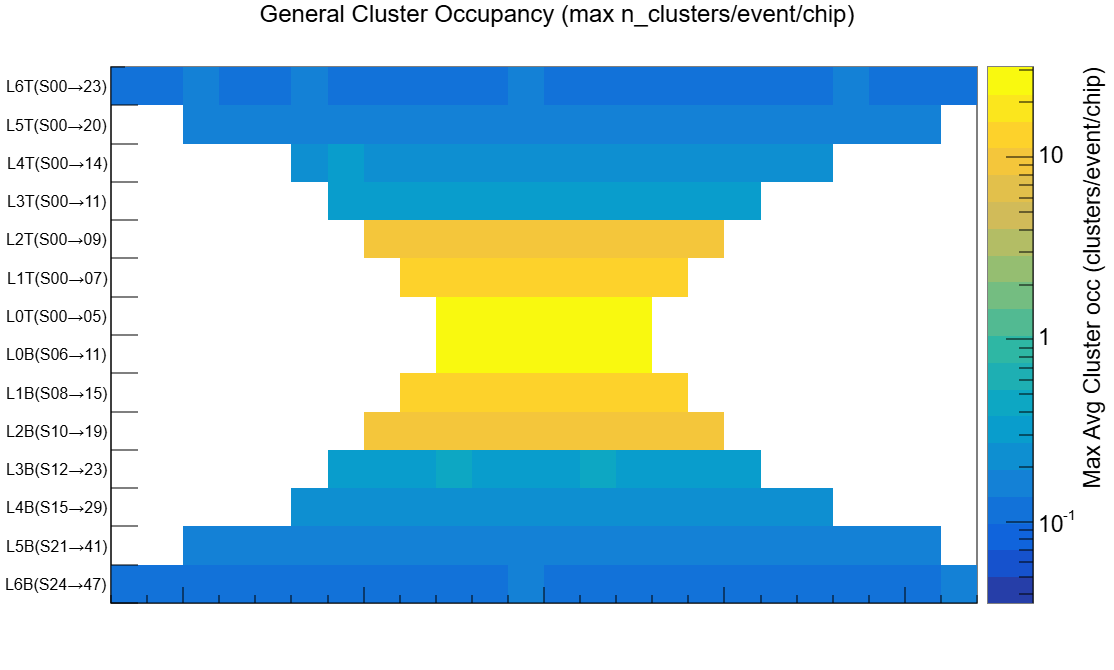}
\caption{Average cluster occupancy. Comparison of synchronous data (left side)  and asynchronous data (right side) reconstruction. Y axis shows layer numbers. X axis shows stave length.
\label{fig:i}}
\end{figure}

\begin{figure}[htbp]
\centering
\includegraphics[width=.45\textwidth]{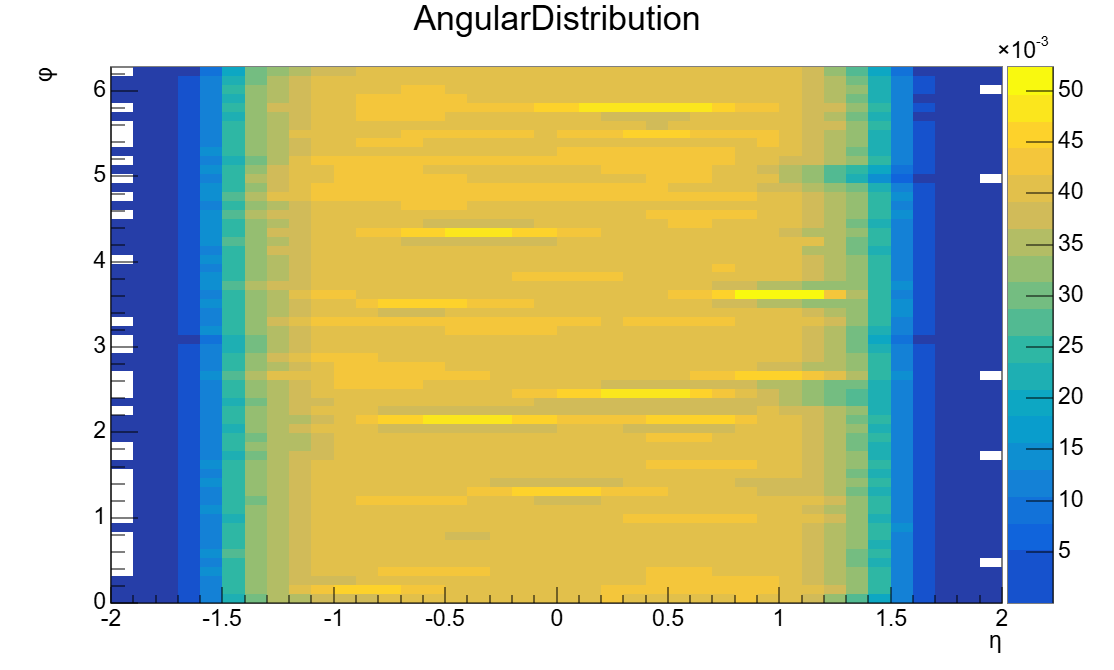}
\qquad
\includegraphics[width=.45\textwidth]{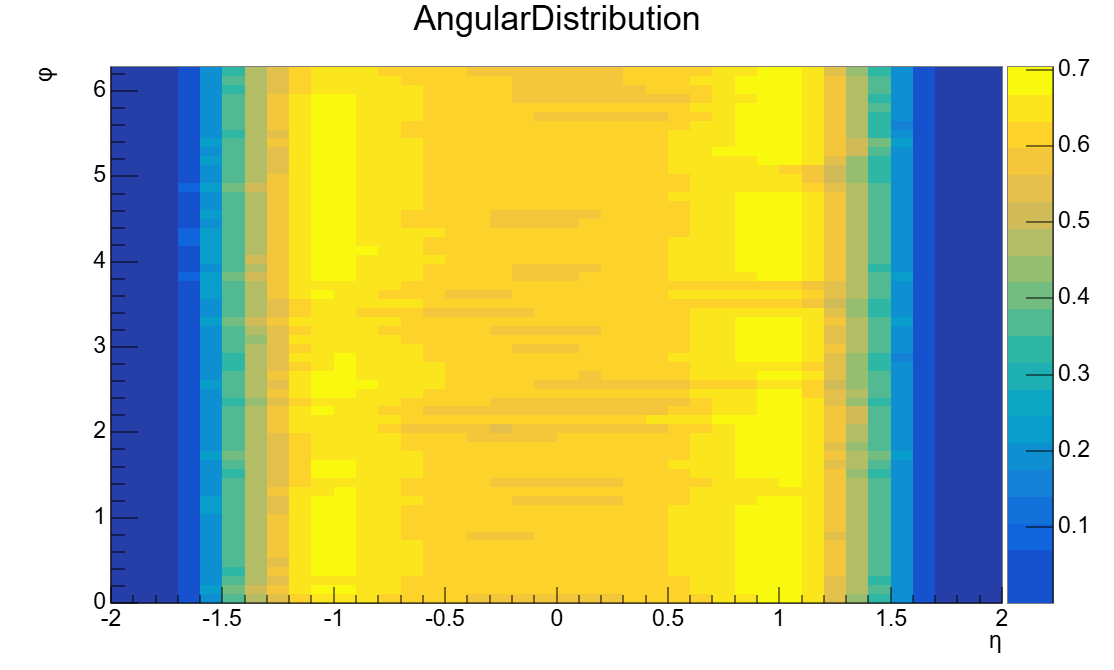}
\caption{Distribution of ITS standalone tracks as a function of azimuthal angle $\varphi$ and pseudorapidity $\eta$. Comparison of synchronous data (left side)  and asynchronous data (right side) reconstruction. 
\label{fig:i}}
\end{figure}

The color scale in Figure 2 shows the maximum of cluster occupancy among all chips or lanes in a stave. It is per each stave in IB, while it shows on lanes in OB. One lane hosts three chips in the ITS inner barrel or seven chips in the ITS outer barrel. The occupancy is calculated as the number of clusters per readout frame per chip.
The plot illustrates how the detector occupancy changes with the distance from the interaction region. The occupancy depends on the interaction rate and framing rate. 
A role of the occupancy plot in the QC is the checking of the uniformity to identify problematic stave with less occupancy or to spot some beam effects. 

Figure 3 shows a distribution of tracks in the ITS2 as a function of azimuthal angle and pseudorapidity.
The $z$-axis represents the number of tracks normalized by the number of
reconstructed interaction vertices during the run.
Only tracks with one hit in each ITS layers and transverse momentum
above 150 MeV/$c$ are shown in the plot, demonstrating that high-quality tracks have uniform acceptance across the ITS2 detector. 
The quality criteria that can be judged from Fig.3 is
an identification of regions with less number of tracks that can fix some issues with hardware in that area.
QC aims for uniform performance across whole azimuth and number of tracks or length of tracks help us to spot more fundamental issues with the detector.

\section{Conclusion}

The new MAPS-based ITS2 has been designed and constructed with the primary goal of enhancing the ALICE track and vertex reconstruction capabilities, in particular at low transverse momentum.
The Data Quality Control software is available to monitor the detector and synchronously/asynchronously check the quality of the data.
The detector has been successfully operated in pp and Pb--Pb collisions with a very low noise level and stable pixel charge threshold.

\acknowledgments

This research was supported by the Ministry of Education, Youth, and Sports of the Czech Republic, project LM2023040.

\end{document}